\documentclass[pdftex,aps,prl,amsmath,amsfonts,amssymb,superscriptaddress,twocolumn,showpacs]{revtex4}

\usepackage[pdftex]{graphicx}
\usepackage{wasysym}
\usepackage{setspace}
\usepackage{color}
\usepackage{nicefrac}

\newcommand{\figref}[1]{Fig.~\ref{fig:#1}}
\newcommand{\Figref}[1]{Figure~\ref{fig:#1}}

\renewcommand{\eqref}[1]{Eq.~(\ref{eq:#1})}
\newcommand{\eqnumref}[1]{(\ref{eq:#1})}

\def\a{s}
\def\b{s}
\newcommand{\add}[1]{\if\a\b{{\color{red} #1}}\else{#1}\fi}
\newcommand{\comm}[1]{\if\a\b{{\color{blue}\{\small \sc #1\}}}\else{}\fi}
\newcommand{\del}[1]{{\if\a\b{{\color{magenta}[[#1]]}}\else{}\fi}}

\begin{document}

\title{Achieving a Strongly Temperature-Dependent Casimir Effect}

\author{Alejandro W. Rodriguez}
\affiliation{Department of Physics,
Massachusetts Institute of Technology, Cambridge, MA 02139}
\author{David Woolf}
\affiliation{Department of Applied Physics,
Harvard University, Cambridge, MA 02139}
\author{Alexander P. McCauley}
\affiliation{Department of Physics,
Massachusetts Institute of Technology, Cambridge, MA 02139}
\author{Federico Capasso}
\affiliation{Department of Applied Physics,
Harvard University, Cambridge, MA 02139}
\author{John D. Joannopoulos}
\affiliation{Department of Physics,
Massachusetts Institute of Technology, Cambridge, MA 02139}
\author{Steven G. Johnson}
\affiliation{Department of Mathematics,
Massachusetts Institute of Technology, Cambridge, MA 02139}

\begin{abstract}
  We propose a method of achieving large temperature $T$ sensitivity
  in the Casimir force that involves measuring the stable separation
  between dielectric objects immersed in a fluid. We study the Casimir
  force between slabs and spheres using realistic material models, and
  find large $> 2$nm/K variations in their stable separations
  (hundreds of nanometers) near room temperature. In addition, we
  analyze the effects of Brownian motion on suspended objects, and
  show that the average separation is also sensitive to changes in
  $T$. Finally, this approach also leads to rich qualitative
  phenomena, such as irreversible transitions, from suspension to
  stiction, as $T$ is varied.
\end{abstract}

\maketitle

Casimir forces between macroscopic objects arise from thermodynamic
electromangetic fluctuations, which persist even in the limit of zero
temperature due to quantum-mechanical effects (the Bose--Einstein
distribution of the photon fluctuations)~\cite{Lifshitz80}.  In most
vacuum-separated geometries, such as parallel metal plates, the force
is attractive and decaying as a function of plate--plate
separation~\cite{Lifshitz80}, becoming readily observable at micron
and submicron separations.  For a nonzero temperature $T$, the force
is predicted to change as a consequence of the changing photon thermal
distribution, but this change is typically negligible near room
temperature and submicron separations~\cite{milton04, Hoye05} and is
only a few percent for $\sim 100$~K changes in $T$ at 1--2\,$\mu$m
separations where Casimir forces are barely
observable~\cite{Lamoreaux97, milton04, Hoye05}.  Therefore, despite
theoretical interest in these $T$ effects~\cite{milton04, Bentsen05},
it has proven difficult for experiments to unambiguously observe $T$
corrections to the Casimir force~\cite{Lamoreaux97}.  Other attempts
to measure $T$ Casimir corrections have focused on nonequilibrium
situations that differ conceptually from forces due purely to
equilibrium fluctuations~\cite{Obrecht07}. A clear experimental
verification of a $T$ Casimir correction would be important in order
to further validate the foundation of Lifshitz theory for Casimir
effects~\cite{milton04, Hoye05}.


In this letter, we propose a method for obtaining strongly
temperature-dependent Casimir effects by exploiting geometries
involving fluid-separated dielectric objects (with separations in the
hundreds of nanometers). In fluid-separated geometries, the Casimir
force can be repulsive~\cite{Dzyaloshinskii61, Lifshitz80}, and can
even lead to stable suspensions of objects due to force-sign
transitions from material dispersion~\cite{Rodriguez08:PRL,
  RodriguezMc10:PRL} or gravity~\cite{McCauleyRo10:PRA,
  RodriguezMc10:PRL}.  We show that, by a proper choice of
materials/geometries, this stable separation $d$ can depend
dramatically on $T$ (2~nm/K is easily obtainable), and there can even
be transitions where $d$ jumps discontinously at some
$T$. Essentially, a stable separation arises from a delicate
cancellation of attractive and repulsive contributions to the force
from fluctuations at different frequencies, and this cancellation is
easily altered or upset by the $T$ corrections. This appears to be the
first prediction of a strong $T$-dependent Casimir phenomenon at
submicron separations where Casimir effects are most easily observed.
We present the phenomenon in simple parallel-plate geometries, but we
believe that the basic idea should extend to many other geometries and
materials combinations that have yet to be explored.  Finally, we also
point out that the same systems that are strongly $T$-dependent can
also be very sensitive to the precise details of the material
dispersion at low frequencies, a property that we plan to exploit in
the future.



The Casimir force between two bodies is a combination of fluctuations
at all frequencies $\omega$, and at $T=0$ can be expressed as an
integral $F(0) = \int_0^\infty f(\xi) d\xi$ over imaginary frequencies
$\omega = i\xi$~\cite{Lifshitz80}.  The contributions $f(\xi)$ from
each imaginary frequency are a complicated function of the geometry
and materials, but they can be computed in a variety of ways, such as
mean stress tensors and the fluctuation--dissipation
theorem~\cite{Dzyaloshinskii61} (valid in fluids for subtle
reasons~\cite{Pitaevski06, Milton10}) or via the Casimir
energy~\cite{Milton10}.  At a finite $T$, this integral is replaced by
a sum over ``Matsubara frequencies'' $2\pi n k T/\hbar = n \xi_T$ for
integers $n$:
\begin{equation}
  F(T) = \frac{2\pi kT}{\hbar} \left[ \frac{f(0^+)}{2} +
    \sum_{n=1}^\infty f\left(\frac{2\pi kT}{\hbar} n\right) \right].
\label{eq:F-T}
\end{equation}
Physically, this arises as a consequence of the $\coth(\hbar \omega /
2kT)$ Bose--Einstein distribution of fluctuations at real
frequencies---when one performs a contour integration in the
upper-half complex-$\omega$ plane, the residues of the $\coth$ poles
at $\hbar \omega / 2kT = n i \pi$ lead to the
summation~\cite{Lamoreaux97}.  Mathematically, \eqref{F-T} corresponds
exactly (including the $\nicefrac{1}{2}$ factor for the zero-frequency
contribution) to a trapezoidal-rule approximation to the $F(0)$
integral, which allows one to use the well-known convergence
properties of the trapezoidal rule~\cite{boyd01:book} to understand
the magnitude of the $T$ correction.  In particular, the difference
between the trapezoidal rule and the exact integral scales as $O(T^2)$
for smooth $f(\xi)$ with nonzero derivative $f'(0)$~\cite{milton04}
(typical for Casimir forces between metals~\cite{Hoye05}).  More
specifically, $f(\xi)$ is exponentially decaying with a decay-length
$2\pi c/a$ for some characteristic lengthscale $a$ (e.g. the
separation or size of the participating objects), while the discrete
sum of \eqref{F-T} corresponds to a lengthscale given by the Matsubara
wavelength $\lambda_T = 2\pi c/\xi_T = \hbar c/kT$, in which case one
would expect the $T$ correction to scale as $O(\lambda_T^2/a^2)$.
Unfortunately, at $T=300$~K, $\lambda_T \approx 7.6\,\mu$m, which is
why the $T$ corrections are typically so small unless $a >
1\,\mu$m~\cite{milton04}.  We cannot change the smoothness of
$f(\omega)$ since it arises from the analyticity of the classical
electromagnetic Green's function in the upper-half complex-$\omega$
plane~\cite{Lifshitz80}, so the only way to obtain a larger $T$
correction is to introduce a longer lengthscale $\Lambda$ into the
problem that dominates over other lengthscales such as the separation
$a$.  One way of achieving this is to make the $f(\xi)$ integrand
\emph{oscillatory} with an oscillation period $\Delta\xi \sim 2\pi
c/\Lambda$ that is much shorter than the decay length $\sim 2\pi c/a$.
Intuitively, discretizing an oscillatory integral induces much larger
discretization effects than for a non-oscillatory integral, and this
intuition can be formallized by a Fourier analysis of the convergence
rate of the trapezoidal rule~\cite{boyd01:book}.  The question then
becomes: how does one obtain an oscillatory Casimir integral?

One way to obtain oscillatory frequency contributions to the Casimir
force is to employ a system where there are combinations of attractive
and repulsive contributions. In particular, it is well known that the
sign of $f(\xi)$ between two dielectric objects emebedded in a fluid
depends on the ordering of their dielectric functions at
$\xi$~\cite{Dzyaloshinskii61, Munday08}:
\begin{equation}
  \mathrm{sgn}(f(\xi)) =
  \begin{cases}
    -1, & \varepsilon_1(i\xi) <  \varepsilon_{fluid}(i\xi) < \varepsilon_2(i\xi) \\
    1, & \mathrm{otherwise},
  \end{cases}\    
\label{eq:eps-stab}
\end{equation}
where $+/-$ denotes an attractive/repulsive force. Since the Casimir
force depends on the dielectric response of the participating objects
over a wide range of $\xi$, from $\xi=0$ all the way to $\xi \sim 2\pi
c/ a$ (where $a$ is a characteristic lengthscale), the sign and
magnitude of the total force at any given separation can be changed by
a proper choice of material dispersion, leading to the possibility of
obtaining Casimir equilibria between objects at multiple
separations. This idea was recently exploited to demonstrate the
possibility of obtaining stable nontouching configurations of
dielectric objects amenable to
experiments~\cite{RodriguezMc10:PRL}. In this paper, for the purpose
of achieving a strong $T$-dependence at short (submicron) separations,
we search for materials or geometries with dielectric crossings
ocurring at sufficiently small $\xi = 2\pi c / \Lambda \sim \xi_T$,
close to the room-temperature Matsubara-frequency scale $\xi_T$.

\begin{figure}[t]
\includegraphics[width=0.9\columnwidth]{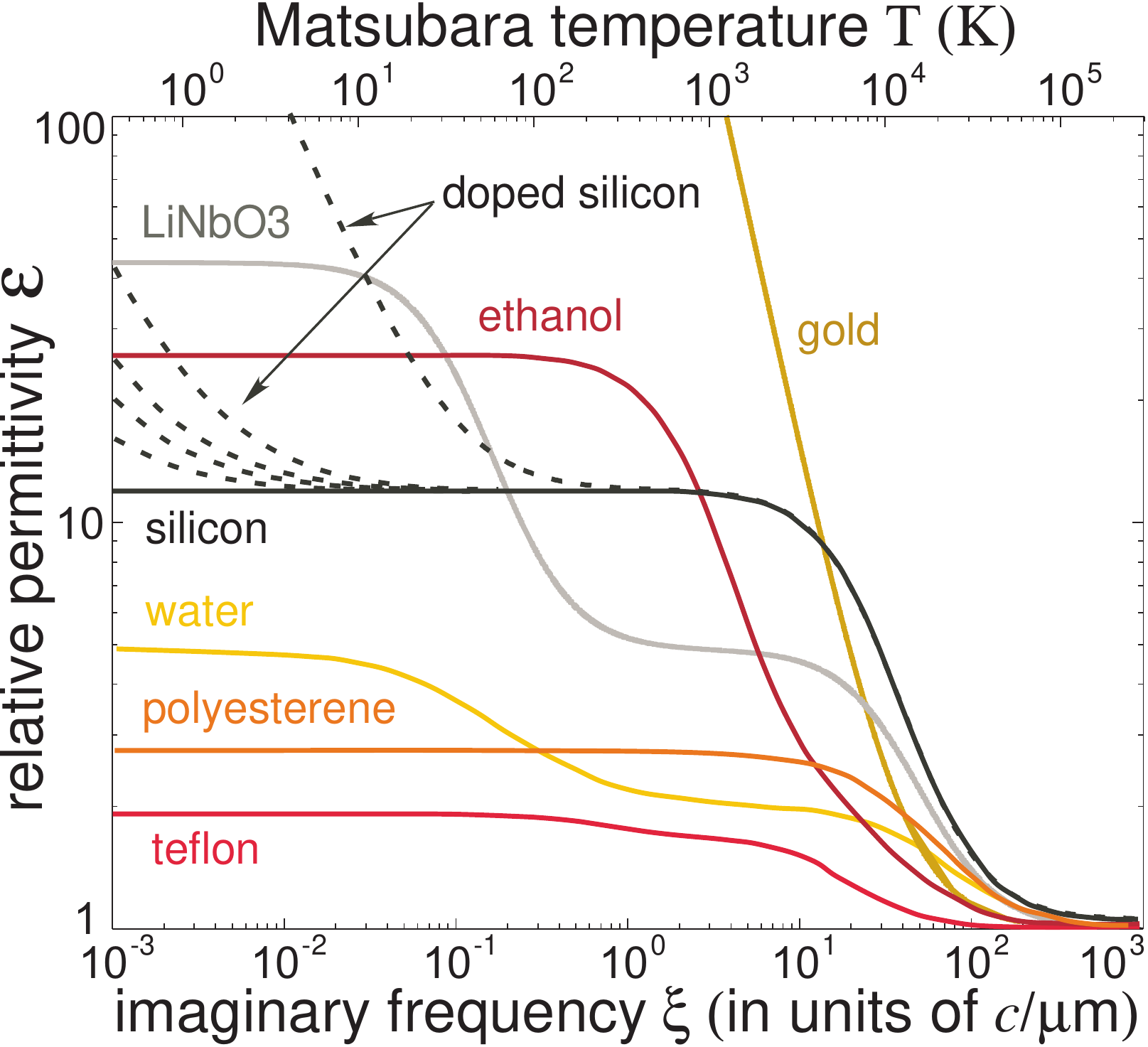}
\caption{Relative permittivity $\varepsilon(i\xi)$ of various
  materials as a function of imaginary frequency $\xi$ (in units of
  $c/\mu$m) or ``Matsubara'' temperature $T = \hbar \xi / 2\pi
  k_\mathrm{B}$. Doped silicon corresponds (bottom to top) to doping
  density $\rho_d = \{1, 3, 5, 10 ,10^2\} \times 10^{16}$, modeled via
  an empirical Drude model~\cite{Duraffourg06}, as is
  gold~\cite{Bergstrom97}. Water, polystyrene, ethanol, teflon, and
  lithium niobate are all modeled via standard Lorentz-oscillator
  models~\cite{Mahanty76}.}
\label{fig:epsilon}
\end{figure}

\begin{figure}[t]
\includegraphics[width=0.9\columnwidth]{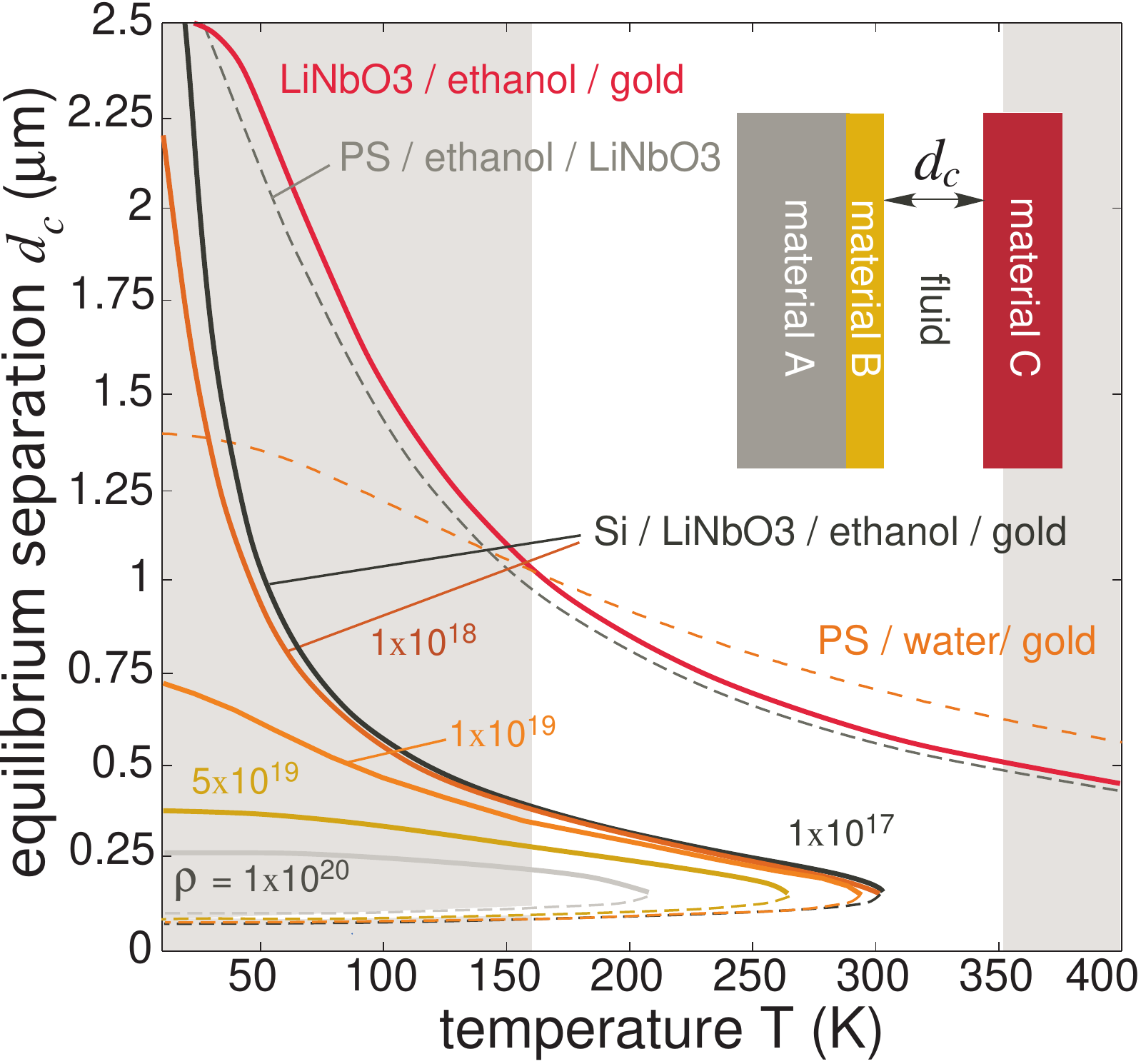}
\caption{Equilibrium separation $d_c$ (in units of $\mu$m) as a
  function of temperature $T$ (in Kelvin), for a geometry consisting
  of fluid-separated semi-infinite slabs (no gravity). The various
  curves correspond to $d_c$ for various material
  combinations. Solid/dashed lines correspond to stable/unstable
  equilibria, and shaded regions are $T$ where ethanol is non-liquid
  at 1~atm~\cite{Friend94}. Doped-silicon is plotted for various
  doping densities $\rho_d = \{1,10,100,500,10^3\}\times 10^{17}$.}
\label{fig:dvsT}
\end{figure}

To begin with, we compute the Casimir force between semi-infinite
slabs, computed via a generalization of the Lifshitz
formula~\cite{Tomas02} that can handle multi-layer dielectric objects,
with relative permittivities $\varepsilon$ plotted in \figref{epsilon}
as a function of imaginary frequency $i\xi$ (bottom axis) or
``Matsubara temperature'' $T = \hbar \xi / 2 \pi k$ (top
axis). \Figref{dvsT} shows the equilibrium separation $d_c$ (in units
of $\mu$m) as a function of temperature $T \in (0,400)$~K (in Kelvin)
for some of the material combinations (solid/dashed lines correspond
to stable/unstable equilibria), and demonstrates various degrees
of~$T$ sensitivity. The previously studied~\cite{RodriguezMc10:PRL}
material combination of teflon/ethanol/silicon (data not shown) shows
very little $T$-dependence: $d_c$ varies $< 1\%$ over $400$~K. More
dramatic behavior is obtained for lithium niobate (LiNbO3) or doped
silicon (doping density $\rho_d = \{1, 10, 100, 500, 10^3\} \times
10^{17}$), whose low-$\xi$ $\varepsilon$ crossings with ethanol lead
to the desired oscillatory $f(\xi)$ in~\eqnumref{F-T}: the
stable-equilibrium separation $d^{(s)}_c$ for both cases decreases by
$> 2\,\mu$m over $400$~K, and $\frac{d}{dT} d^{(s)}_c \approx -2$~nm/K
near $T=300$~K (where $d^{(s)}_c \sim $ 300--700~nm). The sign of
$\frac{d}{dT} d^{(s)}_c$ comes from the increasing domination of the
repulsive small-$\xi$ (large-separation) contributions to $f(\xi)$
as~$T$ increases.  Varying the silicon doping density dramatically
changes the $T$ dependence because it tunes the low-$\xi$
silicon/ethanol $\varepsilon$ crossing in \figref{epsilon}.
Doped-silicon exhibits another interesting behavior: $d_c$ disappears
at a critical temperature $T_c$ (determined by $\rho_d$) due to a
bifurcation between the stable/unstable equilibria. ($T_c$ can be
tuned not only by changing $\rho_d$ but also by changing the LiNbO3
layer-thickness, here $\sim 50$nm.) Experimentally, such a bifurcation
yields an \emph{irreversible} transition from suspension ($T < T_c$)
to stiction ($T > T_c$). \Figref{dvsT} also shows a small sample of
the many other material possibilities.  The shaded regions in
\figref{dvsT} correspond to $T$ above the boiling point (320~K) or
below the freezing point (159~K) of ethanol at 1~atm~\cite{Friend94}.

\begin{figure}[t]
\includegraphics[width=0.9\columnwidth]{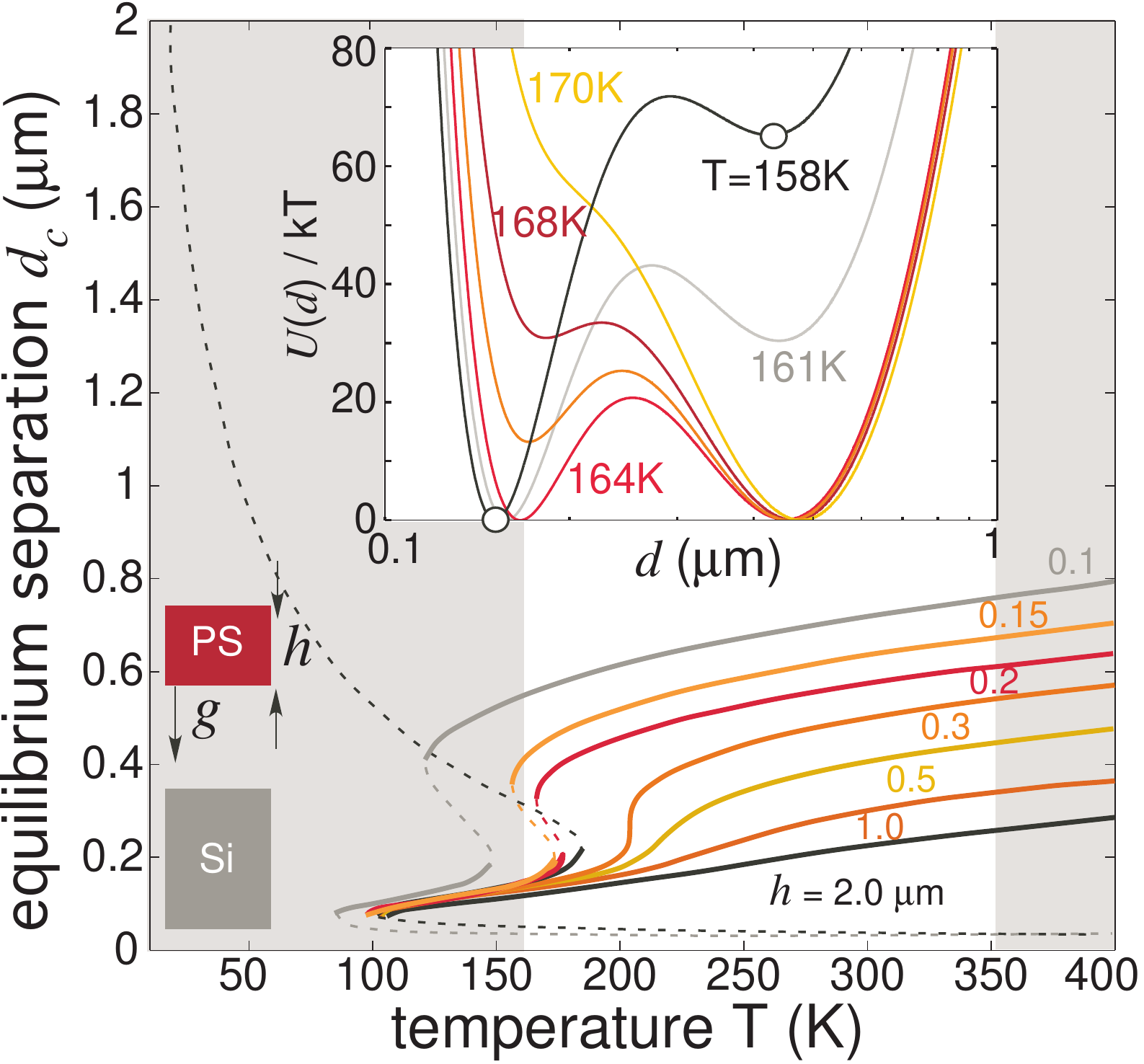}
\caption{Equilibrium position $d_c$ (in units of $\mu$m) of a
  semi-infinite polystyrene (PS) slab immersed in ethanol (shaded~$T=$
  non-liquid) and suspended against gravity by a repulsive Casimir
  force exerted by a doped-silicon (Si) slab. The solid/dashed lines
  correspond to stable/unstable $d_c$, and each color represents a
  different value of PS slab-thickness $h$ (in units of $\mu$m). The
  inset shows the magnitude of the total energy $U_T(d)$ (in units of
  $k_\mathrm{B} T$) as a function of $d$ for $h=150$~nm, at various
  $T$.}
\label{fig:plate-g}
\end{figure}

The inclusion of gravity/buoyancy introduces another force into the
system and leads to the possibility of additional phenomena, such as
additional stable equilibria due to gravity/Casimir
competition~\cite{RodriguezMc10:PRL}.  For example, \figref{plate-g}
shows the equilibrium separations $d_c$ of a polystyrene~(PS) slab of
thickness $h$ in ethanol above a semi-infinite doped-silicon slab
($\rho_d = 1.1\times 10^{15}$), including gravity (mass density
$\rho_\mathrm{PS} - \rho_\mathrm{ethanol} =
0.264$~g/$\mathrm{cm}^3$~\cite{Friend94}).  As in \figref{dvsT}, $d_c$
varies dramatically with $T$: $\frac{d}{dT} d_c \approx 1.2$~nm/K near
$T=300$~K. Gravity becomes increasingly important as~$h$ grows:
compared to $h=0$ (leftmost line), it creates an additional stable
equilibrium (solid lines) at large $d_c$ (hundreds of~nm) where the
downward gravity dominates. With gravity, there are three
stable/unstable bifurcations instead of two, leading to three critical
temperatures where qualitative transitions occur: $T_g$ refers to the
temperature of the topmost bifurcation, created by gravity, and the
other two temperatures are labeled $T_1$ ($\approx 100$~K) and $T_2$
($\approx 180$~K).  If $T_g < T_1$ ($h < 40$~nm), there exists an
irreversible transition from suspension to stiction as $T$ is
decreased below $T_g$.  If $T_1 < T_g < T_2$, there are two
irreversible transitions from suspension to suspension (smaller $d_c$)
to stiction as $T$ is lowered from $T>T_g$ to $T < T_1$ starting in
the large-$d_c$ equilibrium. Finally, when $T_g \to T_2$ ($h \approx
300$~nm) the two stable equilibria merge and only the $T_1$
bifurcation remains. Perhaps most interestingly, when this merge
occurs the slope $\frac{d}{dT}d_c$ can be made arbitrarily large but
finite, corresponding to an arbitrarily large (but reversible)
temperature dependence.  For example, $\Delta d_c \approx 130$~nm for
a small change $\Delta T \approx 5$~K around $T_2$, for $h = 300$~nm.

\begin{figure}[t]
\includegraphics[width=0.9\columnwidth]{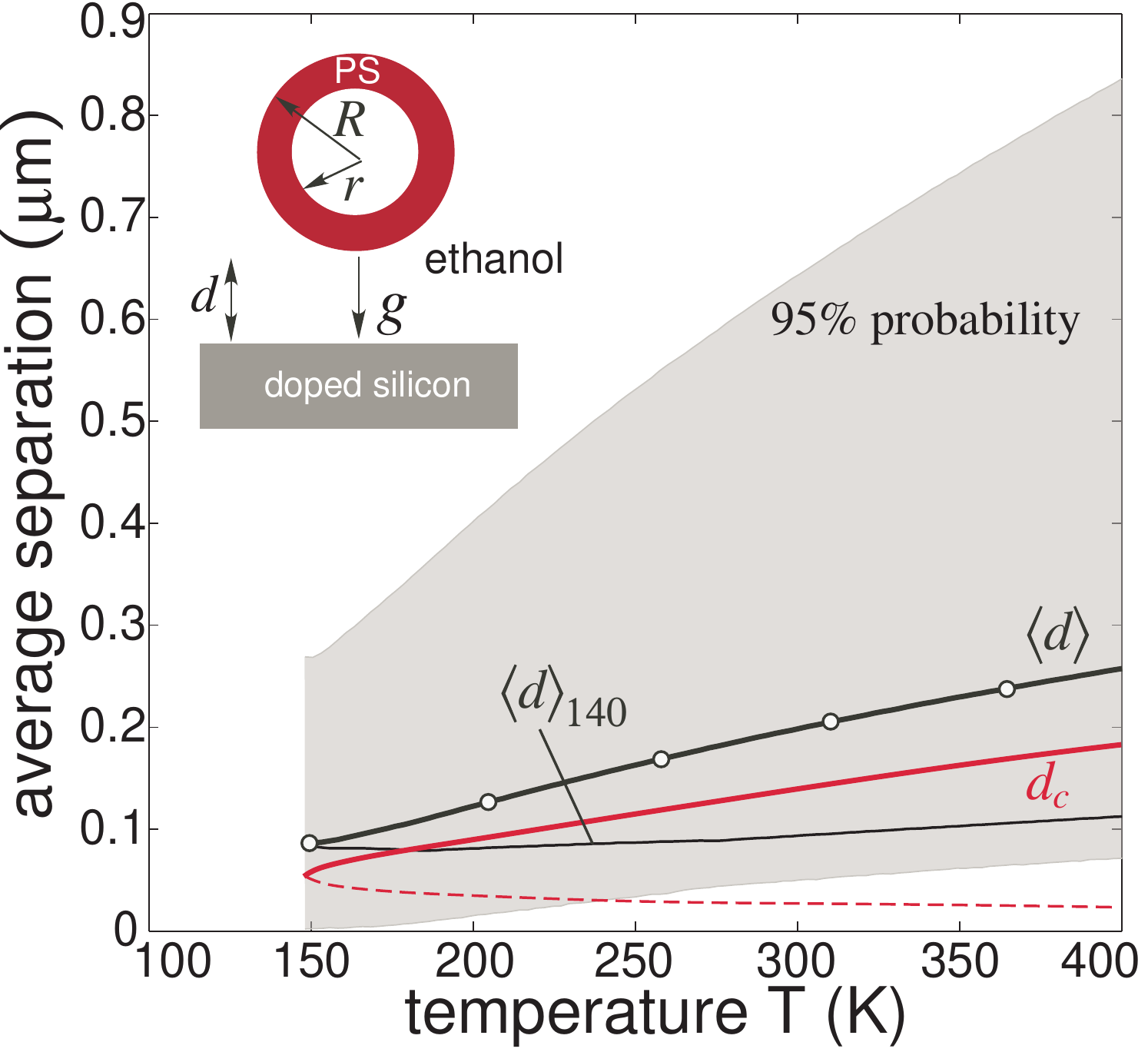}
\caption{Average separation $\langle d \rangle$ (circles) and
  equilibrum separation $d_c$ (red line), in units of $\mu$m,
  vs. temperature $T$ (in Kelvin), for a geometry consisting of a
  fluid-separated hollow PS sphere of inner/outer radius
  $r/R=3.2/5\,\mu$m suspended in ethanol against gravity by a
  doped-silicon slab and subject to Brownian motion.  Shaded region
  indicates where sphere is found with 95\% probability. The thin
  black line is the average $\langle d \rangle_{140}$ if the Casimir
  energy at 140~K is used instead of the true temperature-dependent
  energy landscape.}
\label{fig:davg}
\end{figure}

In a real experiment, the situation is further complicated by Brownian
motion, which will cause the separation to fluctuate around stable
equilibria and will also lead to random transitions between
equilibria~\cite{Risken96}. In the example of \figref{plate-g}, the
attractive interaction at small separations means that there is a
nonzero probability that the slabs will fluctuate past the
unstable-equilibrium energy barrier $\Delta U_T$ into stiction, but
the rate of such a transition decreases proportional to $\exp(-\Delta
U_T/k_\mathrm{B} T)$~\cite{Risken96}---here, assuming a
$50\times50\,\mu\mathrm{m}^2$ PS slab, $\Delta U_T/k_\mathrm{B} T
\approx 10^4$, so the stiction rate is negligible.  The energy
landscape $U_T(d)/k_\mathrm{B}T$ is plotted for several cases in the
inset to \figref{plate-g}: the general prediction of experimental
observations involves a viscosity-damped Langevin
process~\cite{Risken96} that is beyond the scope of this paper to
model, but by choosing $T$ one can make the potential barrier between
the two stable equilibria arbitrarily small and therefore should be
able to reach an experimental regime in which ``hopping'' is
observable.

Alternatively, we consider a simpler example system with only a single
stable equilibrium and a single degree of freedom: a hollow PS sphere
(experimentally available at similar scales~\cite{Wilcox95}), filled
with ethanol, of inner/outer radius $r/R=3.2/5 \,\mu$m suspended in
ethanol above a doped-silicon ($\rho_d = 1.1 \times 10^{15}$)
substrate, shown on the inset of \figref{davg}. (To compute the
Casimir energy in this system, we employ a simple PFA approximation
that is sufficiently accurate for our purpose. Here, for $d \approx
500$~nm, the exact energy is $\approx 85\%$ of the PFA energy.)  For
this example, in \figref{davg} we plot the mean surface--surface
separation $\langle d \rangle \sim \int dz \, z \exp[U_T(z) /
  k_\mathrm{B} T]$ (determined only by the energy landscape and the
Boltzmann distribution~\cite{Risken96}), corresponding to an
experiment averaging $d$ over a long time, along with a confidence
interval (shaded region) indicating the range of $d$ where the
particle is found with 95\% probability. The sphere experiences an
attractive interaction at small separations, but again we find that
the unstable-equilibrium energy barrier is sufficiently large ($\Delta
U / k_\mathrm{B} T \approx 50$) to prevent stiction for $T$ near
300~K. As $T$ varies, two factors affect $\langle d \rangle$: the
$T$-dependence of the Casimir energy $U_T(z)$, and the explicit
$k_\mathrm{B} T$ in the Boltzmann factor.  To distinguish these two
effects, we also plot (thin black line) $\langle d \rangle_{140} \sim
\int dz \, z \exp[U_{140}(z) / k_\mathrm{B} T]$ where the $T=140$~K
(bifurcation point) Casimir energy is used at all temperatures.
Comparing $\langle d \rangle$ with $\langle d \rangle_{140}$, it is
evident that most of the positive-slope $T$ dependence of $\langle d
\rangle$ ($\approx 0.8$~nm/K around 300~K) is due to $U_T$, and
therefore $\langle d \rangle$ offers a direct measure of the
Casimir-energy $T$ dependence.

Experimentally, measuring hundreds of nm changes in separation over
tens or hundreds of Kelvins appears very feasible, perhaps even easier
than traditional measurements of Casimir forces. (In a fluid,
static-charge effects can be neutralized by dissolving electrolytes in
the fluid~\cite{Munday08}, which also have the added benefit of
significantly reducting/increasing the freezing/boiling point of the
fluid~\cite{Friend94}.)  Such temperature-dependent suspensions may
even have practical applications in microfluidics.  We believe that
the examples shown in this letter only scratch the surface of the
possible temperature/dispersion effects that can be obtained in
Casimir-suspension systems. Not only are there many other possible
materials and geometries to explore in the fluid context (along with
more detailed calculation of the Brownian dynamics), and by no means
are the effects shown here the maximum possible, but similar
principles should apply in other systems exhibiting competing
attractive/repulsive Casimir-force contributions.


This work was supported by the Army Research Office through the ISN
under Contract No. W911NF-07-D-0004, by US DOE Grant
No. DE-FG02-97ER25308, and by the Defense Advanced Research Projects
Agency (DARPA) under contract N66001-09-1-2070-DOD.


\end{document}